\documentstyle[epsfig,psfig]{texas}

\begin{document}

\title{EXPLOSIONS DURING GALAXY FORMATION. SCALE-FREE SIMULATIONS}
\author{Hugo MARTEL$^{1}$ and Paul R. SHAPIRO$^{1}$}
\address{(1) Department of Astronomy, University of Texas\\
Austin, TX 78712, USA\\
{\rm Email: hugo@simplicio.as.utexas.edu, shapiro@astro.as.utexas.edu}}

\begin{abstract}
When density fluctuations collapse gravitationally out of the 
expanding cosmological background universe to form galaxies, the secondary
energy release which results can affect their subsequent evolution profoundly.
We focus here on the effects of one form of such
energy release -- explosions, such as might result from the supernovae which
end the lives of the first generation of massive stars to form inside
protogalaxies. We are particularly interested in the consequences of the
nonspherical geometry and continuous infall which are characteristic of 
galaxy formation from realistic initial and boundary conditions.
As an idealized model which serves to illustrate and quantify the importance
of these effects, we study the effect of explosions on the quasi-spherical
objects which form at the intersections of filaments in the plane of
a cosmological pancake, as a result of gravitational instability and
fragmentation of the pancake. We study the formation and evolution of these
``galaxies,'' subject to the explosive injection of energy at their centers,
by numerical gas dynamical simulation in 3D utilizing our new, anisotropic
version of Smoothed Particle Hydrodynamics, Adaptive SPH (``ASPH''), with a 
$\rm P^3M$ gravity solver.
\end{abstract}

\section{INTRODUCTION}

Galaxy formation by gravitational condensation out of the expanding 
cosmological background universe is affected by the complex interplay
of gravitational dynamics of collisionless dark matter and the gas
dynamics of the baryon-electron fluid, including the feedback
on the latter which results when energy is released by the stars
and AGN's that form. Results are presented here of 3D numerical 
gas dynamical simulations of the effect of this energy release, utilizing
our anisotropic Adaptive Smoothed Particle Hydrodynamics (``ASPH'') method.
Current computational limitations make it virtually impossible for numerical
simulations of the initial value problem involving a realistic initial 
spectrum of small-amplitude, Gaussian random noise primordial density
fluctuations to resolve the full range of length and mass scales necessary
to form galaxies and the stars within them at the same time. We choose, 
instead, to focus here on an idealized model of structure formation which
we can hope to resolve more accurately, thereby elucidating some of the 
most important aspects of the problem which may be relevant, as well, to
more realistic initial conditions. We are particularly interested in
studying the effects of nonspherical geometry and continuous cosmological
infall on the problem of explosion-driven blow-out from galaxies during
their formation.

It is well-known that structure 
formation from Gaussian random noise proceeds in a highly anisotropic way,
favoring pancake and filament formation over the formation of quasi-spherical
objects. Our previous work (\cite{ref1},\cite{ref2})
has demonstrated, however, that a cosmological 
pancake, modeled as the nonlinear outcome of a single plane-wave density 
fluctuation, is subject to a linear gravitational instability which
results in the formation of filaments and lumps in the central plane of 
the pancake. The lumps of collisionless dark matter
that form in this way are quasi-spherical and develop
a universal density profile which is reminiscent of the universal
profiles found from N-body simulations of 3D Gaussian noise density 
fluctuations in hierarchical clustering models like the CDM model.
As such, they provide an ideal test-bed for exploring the gas dynamics 
of structure formation and feedback effects without the troublesome complexity
of Gaussian random noise initial conditions. The pancake problem, moreover, is
completely scale-free, once all lengths are expressed in units of
the pancake wavelength $\lambda_p$, time is expressed in terms of the
cosmic scale factor $a$, divided by the scale factor $a_c$ at which the
unperturbed pancake collapses to form density caustics in the dark matter
and shocks in the gas, and the energy release is expressed in units of the
total energy contained in a comoving cube of side $\lambda_p$.
As such, each simulation of galaxy formation via 3D pancake 
gravitational instability 
serves to represent the generic behavior independent of the particular mass
or collapse epoch of the object which forms. In order to preserve this
universality and scale-free character of the problem, we will neglect
radiative cooling in these first calculations. Once the key features of this
scale-free problem are delineated, we will later consider the 
scale-dependent effects of radiative cooling and photoheating.
Among the results we seek to quantify in this way are:

\begin{itemize}

\item The amount of 
energy release required to blow the gas out of a dark matter halo.

\item 
The efficiency for ejecting the fraction of gas which is initially responsible
for receiving the energy release (and, in the case of supernova explosions,
the metallicity associated with the SN ejecta).

\item 
 The distinction between
``blowout,'' in which the energy release results in the escape
of some energy and gas into the surrounding IGM but leaves the
bulk of the gas in the object unaffected, and ``blowaway,'' in which
most or all of the gas is ejected from the dark matter potential well.

\item 
The energy release rate
required to shock-heat the entire IGM by the overlapping effect of energy
release from neighboring objects.

\end{itemize}

\section{PANCAKE INSTABILITY AND FRAGMENTATION AS A TEST-BED
MODEL FOR GALAXY FORMATION}

\subsection{Model and Initial Conditions}

We consider an Einstein-de Sitter universe (density parameter $\Omega_0=1$,
cosmological constant $\lambda_0=0$) with $\Omega_B=0.03$ and $\Omega_X=0.97$
(where $\Omega_B$ and $\Omega_X$ are the contributions of
baryons and dark matter to $\Omega_0$, respectively).
The initial conditions correspond to
the growing mode of a single sinusoidal plane-wave density
fluctuation of wavelength $\lambda_p$ and dimensionless wavevector 
${\bf k}=\hat{\bf x}$ (length unit = $\lambda_p$).
We adjust the initial amplitude $\delta_i$ such that a density caustic forms
in the collisionless dark matter component at scale factor $a=a_c$.

We perturb this system by adding to the initial conditions 
two transverse, plane-wave density fluctuations with equal wavelength
$\lambda_s=\lambda_p$, wavenumbers ${\bf k}_s$ pointing along the
orthogonal vectors
$\hat{\bf y}$ and $\hat{\bf z}$,
and amplitude $\epsilon_y\delta_i$ and $\epsilon_z\delta_i$,
respectively, where $\epsilon_y\ll1$ and $\epsilon_z\ll1$.
We use the notation $S_{1,\epsilon_y,\epsilon_z}$
to designate a pancake perturbed by two such transverse 
perturbation modes. All results presented here refer to the case 
$S_{1,0.2,0.2}$ unless otherwise noted. The presence of the two perturbation
modes will result in the formation of two perpendicular filaments in
the plane of the pancake, with a dense, quasi-spherical cluster at the 
intersection of the filaments.

\subsection{Self-Similar Profiles for Dark Matter Halos}

In 3D, our previous $\rm P^3M$ simulations of collisionless matter involving 
$64^3$ particles and $128^3$ grid cells in a comoving cubic box of side
equal to $\lambda_p$, with gravitational softening parameter 
$\eta=0.3$~grid spacings demonstrated that
the quasi-spherical lumps that form as one of the generic outcomes of 
pancake gravitational instability in 3D, evolve self-similar density 
profiles of universal shape (\cite{ref2},\cite{ref3}).
The universal profile is well-approximated as a power-law
$\rho\propto r^{-2.75}$ over a large range of density, which
flattens somewhat at small radii, a shape 
which is independent of the details of the
initial perturbations to the pancake.
This self-similar profile is similar to the universal 
profile found to fit the results of 3D N-body simulations of the collisionless
dark matter in a CDM model \cite{ref4}. This
suggests that this 3D instability 
of cosmological pancakes which leads generically to the 
formation of such quasi-spherical dark matter halos may be used as an
alternative to the details of the
CDM model with its Gaussian random noise initial density
fluctuations as a test-bed in which to study halo and galaxy formation
further.

To illustrate the universal density profile associated
with this pancake instability, we show in Figure~1
results from \cite{ref2} and \cite{ref3}
for one particular case, $S_{1,0.2,0.2}$, for
various values of the expansion factor, as well as
a summary of results for several different perturbation
modes, at $a/a_c=7$.

\begin{figure}
\centering
\vspace{-5cm}
\psfig{figure=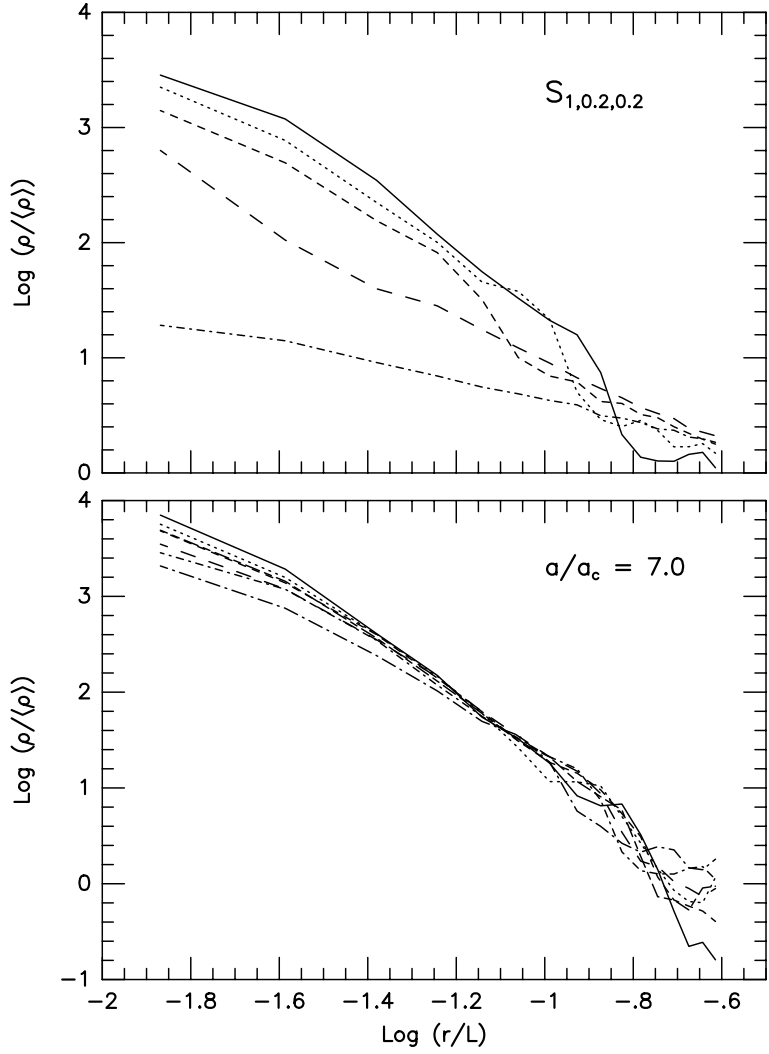,width=15cm}
\caption{Top Panel: Density profiles
(spherically averaged, in units of $\langle\rho\rangle$, the average 
background density) versus radius from halo center (in units of $\lambda_p$)
for mode $S_{1,0.2,0.2}$ at $a/a_c=1$ (dot-short dash), 2 (long dash),
3 (short dash), 4.5 (dotted), and 7 (solid);
Bottom Panel: Density profiles at $a/a_c=7$ for several different modes
of pancake perturbation: $S_{1,1,1}$ (solid), $S_{1,0.5,0.5}$ (dotted),
$S_{1,0.4,0.4}$ (short dash), $S_{1,0.3,0.3}$ (long dash), $S_{1,0.2,0.2}$
(dot-short dash), $S_{1,0.1,0.1}$ (dot-long dash), $S_{1,0.5,0.25}$
(short dash-long dash)(\cite{ref2},\cite{ref3})}\label{fig1}
\end{figure}

\section{THE EFFECT OF EXPLOSIVE ENERGY RELEASE ON GALAXY
FORMATION: BLOW-AWAY AMIDST CONTINUOUS INFALL}

Galaxy formation which leads to star formation can be affected by
feedback when stars evolve to the point of supernova (SN) explosions
and the resulting shock-heating and outward acceleration of interstellar
and intergalactic gas.
Previous attempts to model this effect have typically been along one
of three lines, that which adopts a smooth initial gas distribution in a
galaxy-like, fixed dark matter gravitational potential well 
(e.g. \cite{ref5}), that which considers a single, isolated, but evolving,
density fluctuation (i.e. without merging, infall or the effects
of external tidal forces) (e.g. \cite{ref6}), and that
in which the galaxy forms by condensation out of Gaussian random noise
primordial density fluctuations such as in the CDM model
(e.g. \cite{ref7},\cite{ref8},\cite{ref9},
\cite{ref10},\cite{ref11},\cite{ref12}). 
In the first
case, the computational ability to resolve shocks which propagate
away from the sites of explosive energy release is generally
greater, while the last is perhaps more
realistic in terms of the initial and boundary conditions, but the
resolving of shocks is still generally quite poor.

We compromise here between these two limits by
using the pancake instability problem as the model of galaxy formation in which
to explore the feedback effect of the explosive release of energy by SNe inside
a protogalaxy. This affords some of the benefit of greater ease of the first
approach mentioned above in resolving the explosion-driven shocks which are 
the crucial element in blowing gas out and away. It also provides a 
self-consistent cosmological origin and boundary condition for a protogalaxy
or cluster, including the important effect of anisotropic gravitational 
collapse and continuous infall.

Sharing the initial conditions described above for the 
formation of a dark matter halo via 3D pancake instability mode $S_{1,0.2,0.2}$ 
is an additional component of baryon-electron gas.
We model the explosive release of energy due to
SNe in terms of a single impulsive explosion which may represent a starburst
or the collective effect of multiple SNe. We initiate the explosion
by waiting until the first gas particles at the center of our dark matter
halo reach a density contrast relative to the average background
density, $\rho/\langle\rho\rangle$, exceeding $10^3$, at which point we
suddenly multiply the thermal energy of these particles by a factor $\chi$
and share some of the explosion energy smoothly
amongst their nearest neighbor particles via ASPH kernel smoothing as well.
This occurs at $a/a_c=2.06$. While we have performed a series of
simulations for different values of $\chi$, we shall report here only the
results for two limiting cases: $\chi=0$ (no explosion) and $\chi=10^3$
(blowaway regime). Simulations end at an expansion factor
$a/a_c=3$. (Note: The proper, numerical prescription for depositing the energy
of explosions in the interstellar gas surrounding the explosion site is a
complicated question, dependent as it is on explosion details which are
unresolvable by current numerical treatments. Some recent discussion of this
question is contained in \cite{ref13} and \cite{ref14}, including the question
of the fraction of the energy of a given SN explosion which ends up as
kinetic energy of the SN remnant rather than as thermal energy. However,
since the simulations described here are adiabatic and neglect radiative 
cooling, it is self-consistent for us to deposit the entire
explosion energy initially as thermal energy.)

All our gas dynamical simulations are based upon the new
3D version of our ASPH method 
(\cite{ref15},\cite{ref16},\cite{ref17}),
coupled to a $\rm P^3M$ gravity solver. The simulations reported here use
$32^3$ gas particles, $32^3$ dark matter particles (with unequal particle 
masses, $m_{\rm dark}/m_{\rm gas}=\Omega_X/\Omega_B=32.3$), and
a $\rm P^3M$ grid of $64^3$ cells with softening length $\eta=0.3$ grid 
spacings.

\section{RESULTS}

In the absence of explosion, the simulation produces
two orthogonal
filaments within the pancake central plane at whose intersection is
located a denser, quasi-spherical ball of gas which sits in the 
gravitational potential well of a dense, quasi-spherical dark matter 
halo like that in Figure 1.
In the case with explosion, we found that away
from the central object, the pancake and the filaments within it
are hardly affected. However, gas has been blown out
of the center and some exterior gas which was infalling along
directions perpendicular to the pancake plane has been swept back out, 
as well, some as far as to the outer edge of the box.

In Figure 2, we show a shaded
contour plot of the gas temperature at $a/a_c=3$, in a plane perpendicular 
to the pancake and intersecting the center of the central cluster.
The explosion is confined by the gas in the plane of the pancake
(seen edge-on on this figure)
outside the central object, with the hottest gas at the very center.
The filaments are hardly affected, however, nor is the
shocked pancake gas far from the central object. This edge-on view of the
pancake reveals that a major blow-out has occurred in which
the explosion, led by an outer shock, has propagated all the way to the edge
of the box, half-way to the nearest neighbor pancake, and collided
there with the explosion shock expanding away from the neighboring
pancakes' central object and toward the pancake in this box. The
temperature plot reveals multiple shocks interior to the explosion, 
especially along the symmetry axis of the blow-out. Although the
central object in which the explosion took place was quasi-spherical,
the existence of the pancake plane and of the filaments which intersect
at the location of the central dark halo ensure that a highly anisotropic
explosion results and serves to channel the energy and mass ejection outward
along the symmetry axis.

\begin{figure}
\centering
\vspace{12cm}\hspace{-3cm}
\psfig{figure=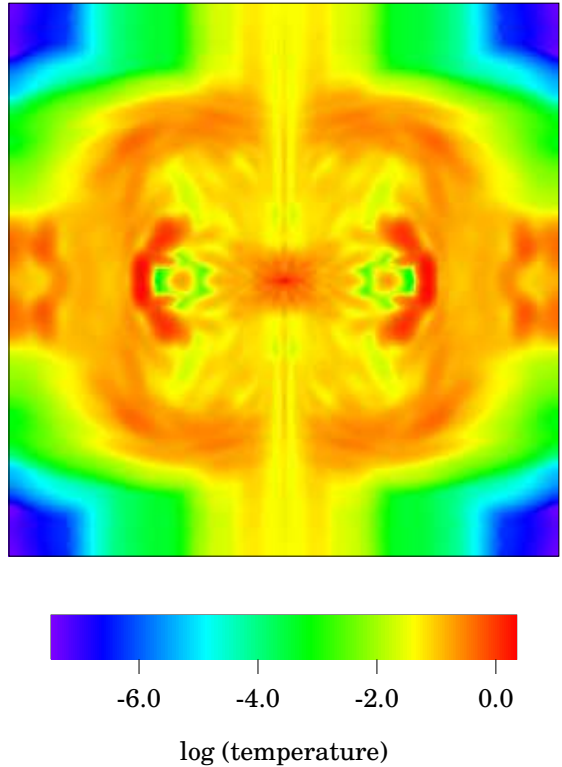,width=9cm}
\vspace{-12cm}
\caption{Dimensionless Gas Temperature 
[i.e. internal energy per gram in units of
$(2401/4)H_0^2\lambda_p^2(a/a_c)^{-4}$] in the plane perpendicular to
the pancake, a cross section view which intersects the
center of the central cluster.}\label{fig2}
\end{figure}

Velocity arrows for the gas particles at $a/a_c=3$ are displayed in 
Figures 3 and 4. Figure~3 shows a thin slice of the computational volume
which contains the pancake central plane (i.e. a top view looking down
on the pancake central plane).
Figure~4 shows a slice perpendicular
to the pancake (that is, the same plane as in Figure~2).
These show that outflow is restricted to the symmetry axis,
while infall continues within the pancake plane and especially
inward along the filaments.

\begin{figure}
\centering
\psfig{figure=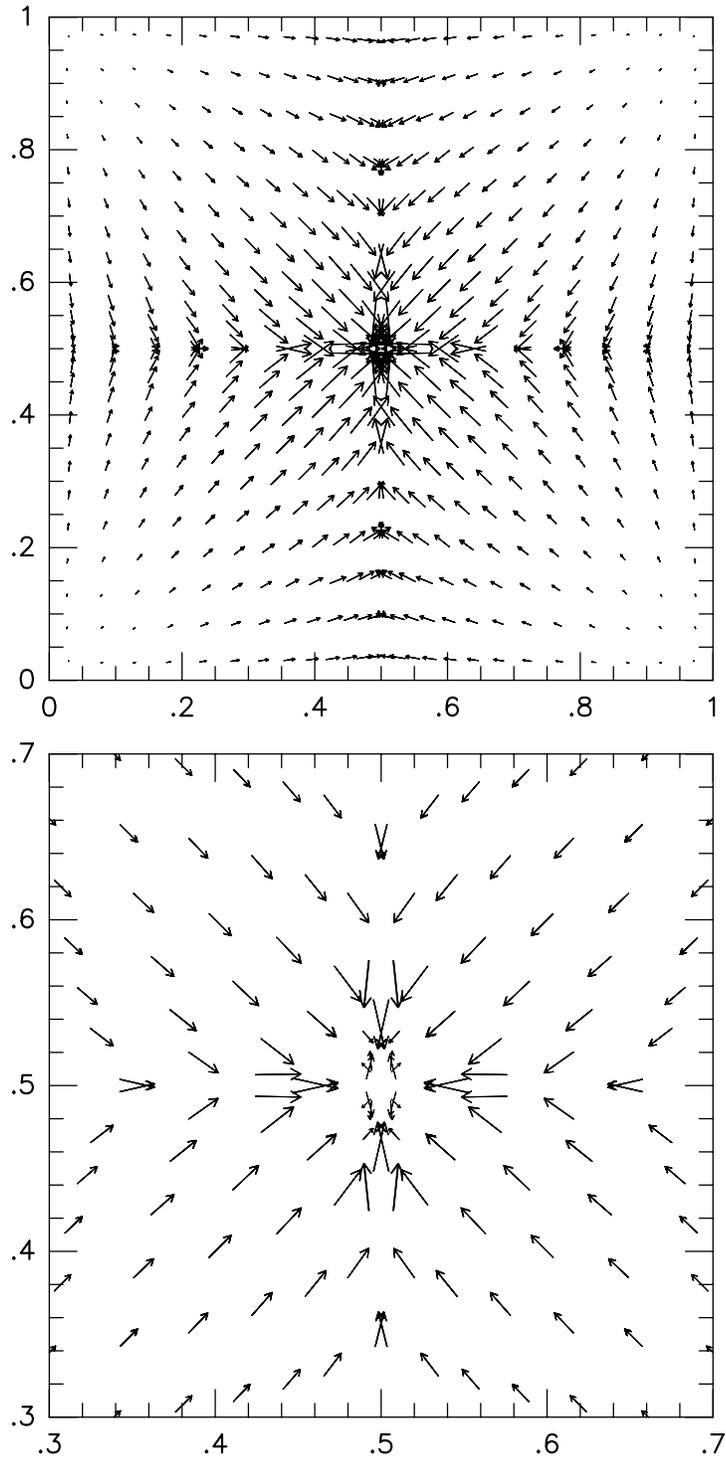,width=15cm}
\caption{Velocity Field at $a/a_c=3$ in the pancake central plane, 
for the case with explosion. 
Each arrow corresponds to a simulation
gas particle in a thin slice containing the pancake symmetry plane.
Top panel: Full image; Bottom panel:
zoom of the central region}\label{fig3}
\end{figure}

\begin{figure}
\centering
\psfig{figure=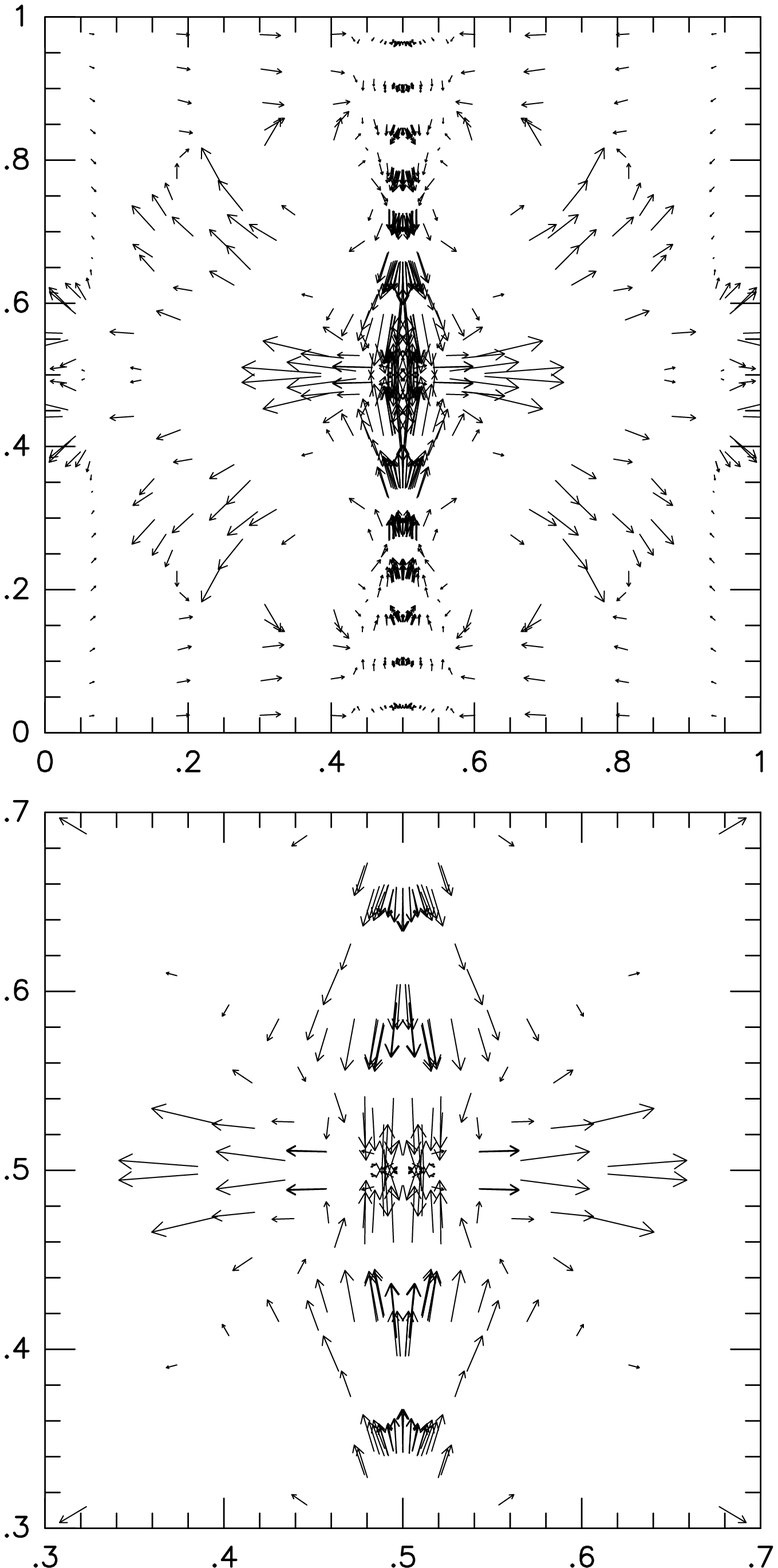,width=15cm}
\caption{Velocity Field at $a/a_c=3$ in the plane perpendicular
to the pancake, 
for the case with explosion. 
Each arrow corresponds to a simulation
gas particle in a thin slice centered on the image plane shown in
Figure 2.
Top panel: Full image; Bottom panel:
zoom of the central region}\label{fig4}
\end{figure}

The effect of the explosion in blowing gas away is
illustrated by Figure 5, in which different particle groups are 
distinguished according to their fate with and without the explosion.
In the top panel, the solid dots show the particles which were the
original recipients of the explosion energy and, by implication,
the metal-enriched SN ejecta,
which were previously located at the very center of the central
dark halo at $a/a_c=2.06$. Gas initially outside this core region
which was within the halo defined according to a mean
overdensity $\rho/\langle\rho\rangle\geq200$ but which did not receive the
initial explosion energy or ejecta directly is shown as open
dots. All of the gas originally inside the halo when the explosion 
occurred has been blown away by $a/a_c=3.0$. The lower panel of Figure 5
shows {\it all} those gas particles which were found to be inside
the central halo with mean overdensity $\rho/\langle\rho\rangle\geq200$
at $a/a_c=3$ in the case with {\it no} explosion but which are {\it not}
inside this overdensity at $a/a_c=3.0$ in the case {\it with} the explosion.

\begin{figure}
\centering
\psfig{figure=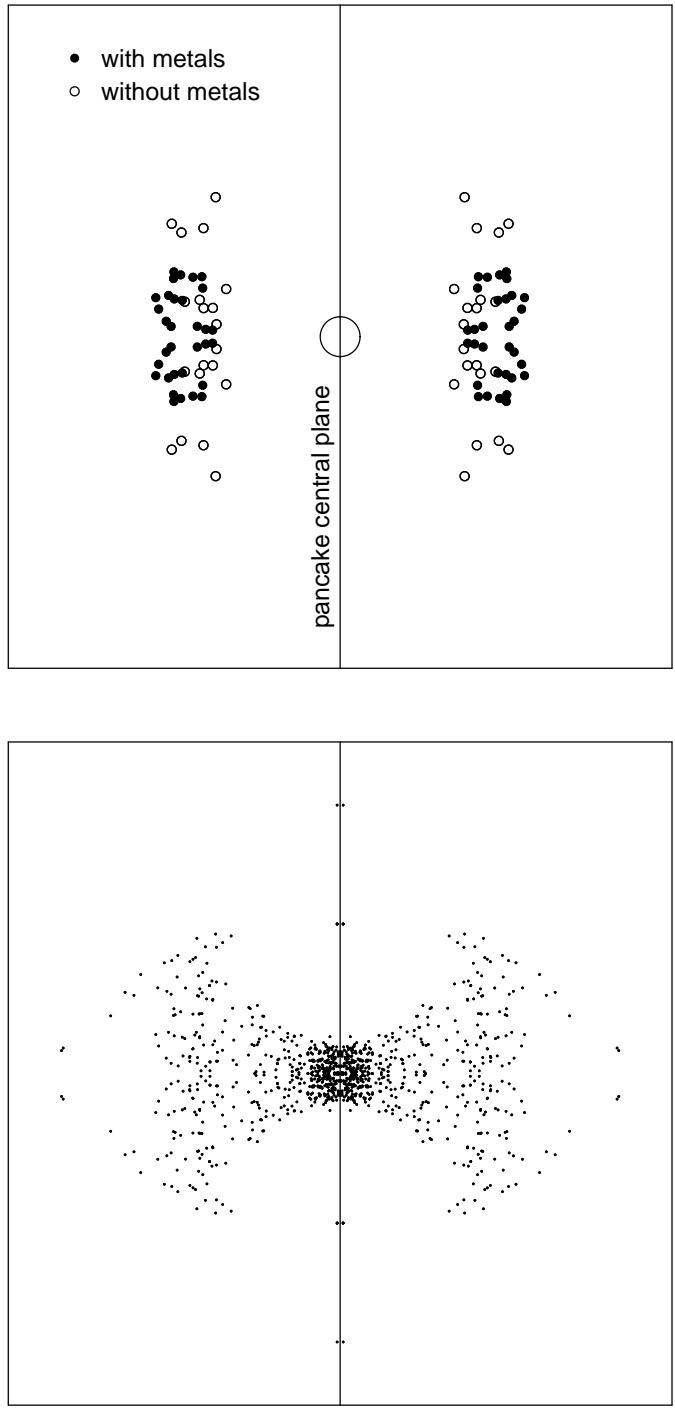,width=15cm}
\caption{Gas particle positions projected onto the image plane of Figure 2.
Case with explosion, at $a/a_c=3$. Top panel: particles that
were located in the central cluster when the explosion occurred. The
filled symbols indicate the particles that were metal-enriched by the 
explosion itself. The line and circle indicate respectively the
location of the pancake plane and the location of the cluster {\it in
the absence of explosion}. 
The radius of the circle is that which enclosed
matter with average density $\rho/\langle\rho\rangle\geq200$ inside
cluster in absence of the explosion.
Bottom panel: particle with density smaller
than $200\langle\rho\rangle$ whose density would be larger
than $200\langle\rho\rangle$ in the absence of explosion}\label{fig5}
\end{figure}

The effect of the explosion on the build-up of
the gas mass of the ``galaxy'' by continuous infall is illustrated by the plot 
in Figure 6 of the collapsed gas fraction in the box, defined as the gas
of overdensity $\rho/\langle\rho\rangle\geq200$, for the cases with and without
the explosion. With no explosion, the central mass grows continuously
from $a/a_c=2$ to $a/a_c=3$ to encompass 10\% of all the mass in
the box. With the explosion, however, all the gas in the central
halo is blown away shortly after the explosion occurs at $a/a_c=2.06$,
but by $a/a_c=2.7$, unobstructed infall within the
pancake plane and along the filaments starts to resupply the central halo with
gas at a significant rate.

\begin{figure}
\centering
\psfig{figure=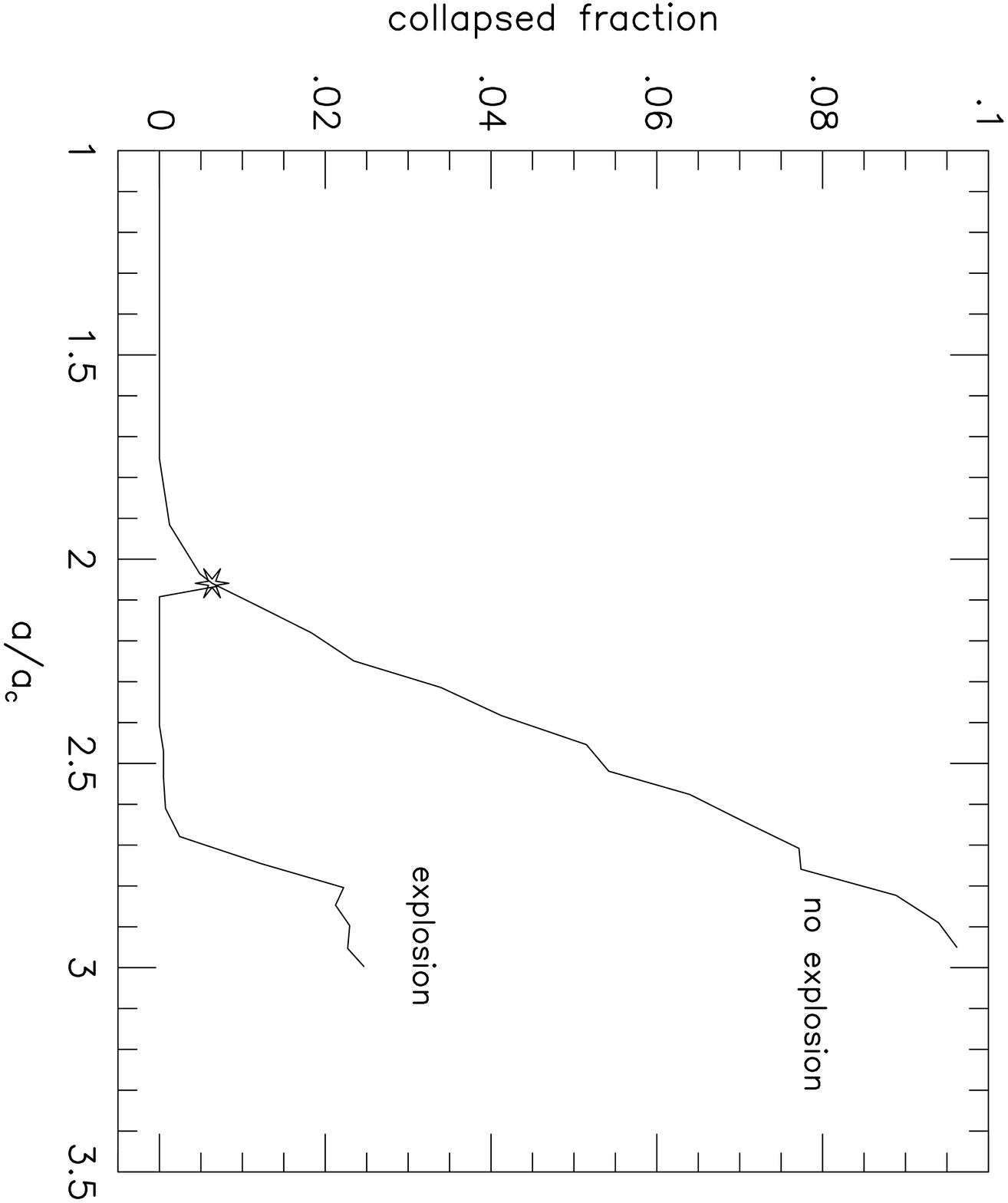,width=9cm}
\caption{Fraction of gas with overdensity $\rho/\langle\rho\rangle\geq200$
versus $a/a_c$}\label{fig6}
\end{figure}

\section{SUMMARY}

\begin{itemize}

\item We find that blow-out and blow-away are generically 
anisotropic events which channel energy and mass loss outward 
preferentially along the symmetry axis of the local pancake and away from the
intersections of filaments in the pancake plane.

\item This means that metal ejection from dwarf galaxies
at high redshift due to explosive energy release is less likely to
pollute the local filaments and pancake in which the dwarf galaxies
reside and more likely to channel the metals away from those denser regions.

\item Despite the complete blow-away of gas 
initially in the dark matter potential well of the ``galaxy''
by the large
explosion simulated here, continuous infall is not completely halted
in the directions away from the preferred direction of blow-out, so infall
partially replenishes the gas which is blown-away.

\end{itemize}

\section*{Acknowledgments}

This work was supported by NASA Grants NAG5-2785, NAG5-7363, and
NAG5-7821, NSF Grants PHY-9800725 and ASC-9504046, and the High Performance 
Computing Facility, University of Texas.

\section*{References}


\end{document}